\documentclass[12pt]{article}
\usepackage{graphics}
\usepackage{amssymb}

\newcommand{\QED}{\mbox{\rule[-1.5pt]{6pt}{10pt}}}
\newtheorem{claim}{Claim}
\newtheorem{theorem}[claim]{Theorem}

\begin{document}

\title{Eigenvalue asymptotics for the Schr{\"o}dinger operator
with a $\delta$-interaction on a punctured surface}
\author{P.~Exner$^{a,b}$ and K.~Yoshitomi$^{c}$}
\date{}
\maketitle
\begin{quote}
{\small \em a) Department of Theoretical Physics, Nuclear Physics
Institute, \\ \phantom{e)x}Academy of Sciences, 25068 \v Re\v z,
Czech Republic \\
 b) Doppler Institute, Czech Technical University,
 B\v{r}ehov{\'a} 7,\\
\phantom{e)x}11519 Prague, Czech Republic \\
 c) Department of Mathematics, Tokyo Metropolitan University,
 \\ \phantom{e)x}Minami-Ohsawa 1-1, Hachioji-shi,
 Tokyo 192-0397, Japan \\
 \rm \phantom{e)x}exner@ujf.cas.cz,
 yositomi@comp.metro-u.ac.jp}
\vspace{8mm}

\noindent MSC numbers: 35J10, 81V99 \\
KEYWORDS: Schr\"odinger
 operators, singular interaction, discrete spectrum, manifolds,
 perturbation
\vspace{5mm}

\noindent {\small {\bf Abstract.} Given $n\geq 2$, we put
$r=\min\{\,i\in\mathbb{N};\: i>n/2\,\}$. Let $\Sigma$ be a
compact, $C^{r}$-smooth surface in $\mathbb{R}^{n}$ which contains
the origin. Let further $\{S_{\epsilon}\}_{0\le\epsilon<\eta}$ be
a family of measurable subsets of $\Sigma$ such that $\sup_{x\in
S_{\epsilon}}|x|= {\mathcal O}(\epsilon)$ as $\epsilon\to 0$. We
derive an asymptotic expansion for the discrete spectrum of the
Schr{\"o}dinger operator $-\Delta -\beta\delta(\cdot-\Sigma
\setminus S_{\epsilon})$ in $L^{2}(\mathbb{R}^{n})$, where $\beta$
is a positive constant, as $\epsilon\to 0$. An analogous result is
given also for geometrically induced bound states due to a
$\delta$ interaction supported by an infinite planar curve. }
\end{quote}
\bigskip


\section{Introduction}

Schr{\"o}dinger operators with $\delta$-interactions supported by
subsets of a lower dimension in the configuration space have been
studied by numerous authors -- see, e.g., \cite{AGHH}--\cite{BEKS}
and references therein. Recently such systems attracted a new
attention as models of ``leaky'' quantum wires and similar
structures; new results have been derived about a
curvature-induced discrete spectrum \cite{EI, EK1} and the
strong-coupling asymptotics \cite{Ex, EK2, EY1, EY2, EY3}.

The purpose of this paper is to discuss another question, namely
how the discrete spectra of such operators behave with respect to
a perturbation of the interaction support. Since the argument we
are going to use can be formulated in any dimension, we consider
here generally $n$-dimensional Schr{\"o}dinger operators, $n\ge
2$, with a $\delta$-interaction supported by a punctured surface.
On the other hand, we restrict our attention to the situation when
the surface codimension is one and the Schr{\"o}dinger operator in
question is defined naturally  by means of the appropriate
quadratic form.

Formally speaking, our result says that up to an error term the
eigenvalue shift resulting from removing an
$\epsilon$-neighbourhood of a surface point is the same as that of
adding a repulsive $\delta$ interaction at this point with the
coupling constant proportional to the puncture ``area''. We will
formulate this claim precisely in Theorem~\ref{main} below for any
sufficiently smooth compact surface in $\mathbb{R}^n$ and prove it
in Section~3. Furthermore, the compactness requirement is not
essential in the argument; in Section~4 we will derive an
analogous asymptotic formula for an infinite planar curve which is
not a straight line but it is asymptotically straight in a
suitable sense.


\section{The main result}

Put $r:=\min\{\,i\in\mathbb{N};\: i>n/2\,\}$. Let $\Sigma$ be a
compact, $C^{r}$-smooth surface in $\mathbb{R}^{n}$ which contains
the origin, $0\in\Sigma$. Let further $\{S_{\epsilon}\}_{0\leq
\epsilon<\eta}$ be a family of subsets of $\Sigma$ which obeys the
following hypotheses:
\begin{description}
 \vspace{-.6ex}
\item{(H.1)} The set $S_{\epsilon}$ is measurable with respect
to the $(n\!-\!1)$-dimensional Lebesgue measure on $\Sigma$ for
any $\epsilon\in [0,\eta)$.
 \vspace{-1.2ex}
\item{(H.2)} $\:{\displaystyle \sup_{x\in S_{\epsilon}}}|x|=
{\mathcal O}(\epsilon)$ as $\,\epsilon\to 0$.
 \vspace{-.6ex}
\end{description}
Next we fix $\beta>0$ and define for $0\leq\epsilon<\eta$ the
quadratic form $q_{\epsilon}$ by
$$ q_{\epsilon}[u,v]:=(\nabla u,\nabla v)
_{L^{2}(\mathbb{R}^{n})}-\beta \int_{\Sigma\setminus S_{\epsilon}}
u(x)\overline{v(x)}\,dS\,, \quad u,v\in H^{1}(\mathbb{R}^{n})\,; $$
it is easily seen to be closed and bounded from below. Let
$H_{\epsilon}$ be the self-adjoint operator associated with
$q_{\epsilon}$. Since $\Sigma\setminus S_{\epsilon}$ is bounded,
we have
 $$\sigma_\mathrm{ess}(H_{\epsilon})=[0,\infty) \quad \mathrm{
and}\quad \sharp \sigma_\mathrm{disc}(H_{0})<\infty\,. $$
By the min-max principle, there exists a unique $\beta^{*}\geq 0$
such that $\sigma_\mathrm{disc}(H_{0})$ is non-empty if
$\beta>\beta^{*}$ while $\sigma_\mathrm{disc}(H_{0})=\emptyset$
for $\beta\leq\beta^{*}$. The critical coupling is
dimension-dependent: a straightforward modification of the usual
Birman-Schwinger argument using \cite[Lemma~2.3]{BEKS} shows that
$\beta^{*}=0$ when $n=2$, while for $n\ge 3$ we have $\beta^{*}>0$
by \cite[Thm 4.2(iii)]{BEKS}. Since our aim is to derive
asymptotic properties of the discrete spectrum, we will assume
throughout that
\begin{description}
 \vspace{-.6ex}
\item{(H.3)} $\:\beta>\beta^{*}.$
\end{description}
Let $N$ be the number of negative eigenvalues of $H_{0}$. Since
$$ 0\leq q_{\epsilon}[u,u]-q_{0}[u,u]\to 0 \quad \mathrm{as} \quad
\epsilon\to 0 \quad\mathrm{for}\quad u\in
H^{1}(\mathbb{R}^{n})\,,$$
there exists $\eta^{\prime}\in (0,\eta)$ such that for
$\epsilon\in (0,\eta^{\prime})$ the operator $H_{\epsilon}$ has
exactly $N$ negative eigenvalues denoted by
$\lambda_{1}(\epsilon)< \lambda_{2}(\epsilon)
\leq\cdots\leq\lambda_{N}(\epsilon)$, and moreover
$$\lambda_{j}(\epsilon)\to\lambda_{j}(0)\quad \mathrm{as}
\quad\epsilon\to 0\quad\mathrm{for}\quad 1\leq j\leq N $$
(see \cite[Chap.~VIII, Thm~3.15]{Ka}). Let $\{\varphi_{j}
(x)\}^{N}_{j=1}$ be an orthonormal system of eigenfunctions of
$H_{0}$ such that $H_{0}\varphi_{j}=\lambda_{j}(0)\varphi_{j}$ for
$1\leq j\leq N$. Pick a sufficiently small $a>0$ so that the set
$\{\,x\in\mathbb{R}^{n}:\: |x|<a\,\}\setminus\Sigma$ consists of
two connected components, which we denote by $B_{\pm}$. We have
$\varphi_{j}\in H^{r}(B_{\pm})$ by the elliptic regularity theorem
 (see \cite[Sec.~10]{A}), because the form domain
 $H^{1}(\mathbb{R}^{n})$ of $q_{0}$ is
locally invariant under tangential translations along the surface
$\Sigma$.
Since $r>n/2$ by assumption, the Sobolev trace
theorem implies that the function $\varphi_{j}$ is continuous on a
$\Sigma$-neighbourhood of the origin. We also note that one can
suppose without loss of generality that $\varphi_{1}(x)>0\,$ in
$\mathbb{R}^{n}$. For a given $\mu\in\sigma_\mathrm{disc}(H_{0})$
we define
 \begin{eqnarray*}
m(\mu) &\!:=\!& \min\{1\leq j\leq N;\: \mu=\lambda_{j}(0)\}\,,
\\ n(\mu) &\!:=\!& \max\{1\leq j\leq N;\:
\mu=\lambda_{j}(0)\}\,, \\ C(\mu) &\!:=\!& \left(\,\varphi_{i}(0)
\overline{\varphi_{j}(0)}\,\right)_{m(\mu)\leq i,j\leq n(\mu)}\,.
 \end{eqnarray*}
Let $s_{m(\mu)}\leq s_{m(\mu)+1}\leq\cdots\leq s_{n(\mu)}$ be the
eigenvalues of the matrix $C(\mu)$. In particular, if
$\mu=\lambda_j(0)$ is a simple eigenvalue of $H_0$, we have
$m(\mu)=n(\mu)=j$ and $s_j= |\varphi_{j}(0)|^2$. Our main result
can be then stated as follows.
\begin{theorem} \label{main}
Adopt the assumptions (H.1)--(H.3). Let $\mu\in
\sigma_\mathrm{disc}(H_{0})$, then the asymptotic formula
$$ \lambda_{j}(\epsilon)=\mu +\beta\,
\mathrm{meas}_{\Sigma}(S_{\epsilon}) s_{j} +o(\epsilon^{n-1})
\quad\mathit{as}\quad \epsilon\to 0 $$
holds for $m(\mu)\leq j\leq n(\mu)$, where
$\mathrm{meas}_{\Sigma}(\cdot)$ stands for the
$(n\!-\!1)$-dimensional Lebesgue measure on $\Sigma$.
\end{theorem}

It should be stressed that our problem involves a singular
perturbation and thus it cannot be reduced to the general asymptotic
perturbation theory of quadratic forms described in
\cite[Sec.~VIII.4]{Ka}. Indeed, we have
$$q_{\epsilon}[u,u]=q_{0}[u,u] +\beta\,
\mathrm{meas}_{\Sigma}(S_{\epsilon}) |u(0)|^{2}+{\mathcal O}
(\epsilon^{n}) \quad\mathrm{as}\quad\epsilon\to 0 $$
for $u\in C^{\infty}_{0}(\mathbb{R}^{n})$ and the quadratic form
$C^{\infty}_{0}(\mathbb{R}^{n}) \owns u\mapsto |u(0)|^{2}\in
\mathbb{R}$ does not extend to a bounded form on
$H^{1}(\mathbb{R}^{n})$, because the set
$$ \left\{\,u\in C^{\infty}_{0}(\mathbb{R}^{n});\:  u=0\quad
\mathrm{in\; a\; neighbourhood\, of\; the\; origin}\, \right\}$$
is dense in $H^{1}(\mathbb{R}^{n})$. We eliminate this difficulty
by using the compactness of the map $H^{1}(\mathbb{R}^{n})\owns
f\mapsto f|_{\Sigma}\in L^{2}(\Sigma)$, which will enable us to
prove Theorem~\ref{main} along the lines of the
asymptotic-perturbation theorem proof.

Let us remark that our functional-analytic argument has a
distinctive advantage over another technique employed in such
situations, usually called the matching of asymptotic expansions
-- see \cite{Il} for a thorough review -- since the latter
typically requires a sort of self-similarity for the perturbation
domains. Our technique needs no assumption of this type.


\section{Proof of Theorem~\ref{main}}

We denote $R(\zeta,\epsilon)= (H_{\epsilon}-\zeta)^{-1}$ for
$\zeta\in\rho(H_{\epsilon})$ and $R(\zeta)=(H_{0}-\zeta)^{-1}$ for
$\zeta\in\rho(H_{0})$. Put $\kappa:= {1\over 2}\,\mathrm{dist}
(\{\mu\}, \sigma(H_{0})\setminus\{\mu\})$. Since
$\lambda_{j}(\cdot)$ is continuous at the origin for $1\leq j\leq
N$, there is an $\eta_{0}\in (0,\eta^{\prime})$ such that
\begin{eqnarray*}
\lefteqn{ \sigma(H_{\epsilon})\cap[\mu-\kappa,\mu+\kappa] =
\sigma(H_{\epsilon})\cap(\mu-\kappa/2,\mu+\kappa/2)}\\ &&
\phantom{AA} = \{\lambda_{m(\mu)}(\epsilon),
\lambda_{m(\mu)+1}(\epsilon),\ldots, \lambda_{n(\mu)}(\epsilon)\}
\end{eqnarray*}
holds if $0<\epsilon \leq\eta_{0}$. Choosing the circle
$C:=\{\,z\in\mathbb{C};\: |z-\mu|={3\over 4} \kappa\,\}$ we put
$$ w_{j}(\zeta,\epsilon):=
R(\zeta,\epsilon)\varphi_{j}-R(\zeta)\varphi_{j} \quad
\mathrm{for}\quad 0<\epsilon \leq\eta_{0}\,,\; \zeta\in C\,.$$
Our first aim is to check that
\begin{equation} \label{wdecay}
\| w_{j}(\zeta,\epsilon)\|_
{H^{1}(\mathbb{R}^{n})}=\mathcal{O}(\epsilon^{(n-1)/2})\quad
\mathrm{as}\quad\epsilon\to 0
\end{equation}
holds uniformly with respect to $\zeta\in C$ for $m(\mu)\leq j\leq
n(\mu)$.
 Notice that there exists a $K_{0}>0$ such that
$$ \Vert u\Vert^{2}_{H^{1}(\mathbb{R}^{n})}\leq
2\left|(q_{\epsilon}-\zeta)[u,u]\right|+K_{0}\Vert
u\Vert^{2}_{L^{2}(\mathbb{R}^{n})}$$
for $\zeta\in C$, $u\in H^{1}(\mathbb{R}^{n})$, and
$0<\epsilon\leq\eta_{0}$. This implies that there exists a
$K_{1}>0$
 such that
$$\Vert R(\zeta,\epsilon)u\Vert_{H^{1}(\mathbb{R}^{n})}\leq K_{1}
\Vert u\Vert_{L^{2}(\mathbb{R}^{n})}$$ for $\zeta\in C$, $u\in
H^{1}(\mathbb{R}^{n})$, and $0<\epsilon\leq\eta_{0}$.
 Moreover, by the Sobolev trace
theorem, there exists a constant $K_{2}>0$ such that
$$ \| u\|_{L^{2}(\Sigma)}\leq K_{2}\| u\|_{H^{1}(\mathbb{R}^{n})}
\quad\mathrm{for}\quad u\in H^{1}(\mathbb{R}^{n})\,.$$
Combining these three estimates we get
\begin{eqnarray}
{}&{}&\| w_{j}(\zeta,\epsilon)\|^{2}_{H^{1}( \mathbb{R}^{n})}\nonumber\\
&\!\leq\!&2\left|(q_{\epsilon}-\zeta) [w_{j}(\zeta,\epsilon),
w_{j}(\zeta,\epsilon)]\right|+K_{0}(w_{j}(\zeta,\epsilon),
w_{j}(\zeta,\epsilon))_{L^{2}( \mathbb{R}^{n})}\nonumber\\
&\!=\!&2\left|-\beta\int_{S_{\epsilon}} R(\zeta)\varphi_{j}
\overline{w_{j}(\zeta,\epsilon)}\,dS\right|
+K_{0}(q_{0}-q_{\epsilon})[R(\zeta)\varphi_{j},R(\overline{\zeta},\epsilon)
w_{j}(\zeta,\epsilon)]\nonumber\\
&\!\leq\!& \beta\Vert
R(\zeta)\varphi_{j}\Vert_{L^{2}(S_{\epsilon})}( 2\Vert
w_{j}(\zeta,\epsilon)\Vert _{L^{2}(S_{\epsilon})}+ K_{0}\Vert
R(\overline{\zeta},\epsilon)w_{j}(\zeta,\epsilon)\Vert
_{L^{2}(S_{\epsilon})})\nonumber\\
&\!=\!&
\frac{4\beta}{3\kappa}\Vert\varphi_{j}\Vert_{L^{2}(S_{\epsilon})}(
2\Vert w_{j}(\zeta,\epsilon)\Vert _{L^{2}(S_{\epsilon})}+
K_{0}\Vert R(\overline{\zeta},\epsilon)w_{j}(\zeta,\epsilon)\Vert
_{L^{2}(S_{\epsilon})})\nonumber\\
&\!\leq\!& \frac{4\beta}{3\kappa} K_{2}(2+K_{0}K_{1})
\Vert\varphi_{j}\Vert_{L^{2}(S_{\epsilon})}
 \| w_{j}(\zeta,\epsilon)\| _{
H^{1}(\mathbb{R}^{n})}\,.\label{west}
\end{eqnarray}
Since $\| \varphi_{j}\|_{ L^{2}(S_{\epsilon})}=\mathcal{O}
(\epsilon^{(n-1)/2})$ as $\epsilon\to 0$ by the assumptions (H.1),
(H.2) and the continuity of $\varphi_{j}|_{\Sigma}$ at the origin,
we arrive at the relation (\ref{wdecay}).

In the next step we are going to demonstrate that the convergence
is in fact slightly faster, namely
\begin{equation} \label{wdecay2}
\sup_{\zeta\in C} \| w_{j}(\zeta,\epsilon)\|_{ H^{1}
(\mathbb{R}^{n})} =o(\epsilon^{(n-1)/2})\quad
\mathrm{as}\quad\epsilon\to 0\,.
\end{equation}
We will proceed by contradiction. Suppose that (\ref{wdecay2})
does not hold; then there would exist a constant $\delta>0$, a
sequence $\{\epsilon_{i}\}^{\infty}_{i=1}\subset (0,\eta_{0})$
which tends to zero, and $\{\zeta_{i}\}^{\infty}_{i=1} \subset C$
such that
\begin{equation} \label{contr}
\epsilon_{i}^{-(n-1)/2}\| w_{j}(\zeta_{i},\epsilon_{i})
\|_{H^{1}(\mathbb{R}^{n})}\geq \delta\quad{\rm for\,\,all} \quad
i\in\mathbb{N}\,.
\end{equation}
Notice that the map $H^{1}(\mathbb{R}^{n})\owns f\mapsto
f|_{\Sigma}\in L^{2}(\Sigma)$ is compact due to the boundedness of
the map $H^{1}(\mathbb{R}^{n})\owns g\mapsto g|_{\Sigma}\in
H^{1/2}(\Sigma)$ and the compactness of the imbedding
$H^{1/2}(\Sigma)\owns h\mapsto h\in L^{2}(\Sigma)$ -- cf. [15,
Chap.~1, Thms 8.3 and 16.1]. Since the two sequences
$$\left\{\epsilon_{i}^{-(n-1)/2}w_{j}(\zeta_{i},\epsilon_{i})
\right\}^{\infty}_{i=1}\quad{\rm and}\quad
\left\{\epsilon_{i}^{-(n-1)/2}R(\overline{\zeta_{i}},\epsilon_{i})
w_{j}(\zeta_{i},\epsilon_{i}) \right\}^{\infty}_{i=1}$$ are
bounded in $H^{1}(\mathbb{R}^{n})$, there is a subsequence
$\{i(k)\}^{\infty}_{k=1}$ of $\{i\}^{\infty}_{i=1}$ such that
$$\left\{
\epsilon_{i(k)}^{-(n-1)/2} w_{j}(\zeta_{i(k)},
\epsilon_{i(k)})\right\}^{\infty}_{k=1} \quad{\rm and}\quad
\left\{ \epsilon_{i(k)}^{-(n-1)/2}
R(\overline{\zeta_{i(k)}},\epsilon_{i(k)}) w_{j}(\zeta_{i(k)},
\epsilon_{i(k)})\right\}^{\infty}_{k=1}
$$
converge in $L^{2} (\Sigma)$. Let us denote
 $$ g:=\lim_{k\to\infty}
\epsilon_{i(k)}^{-(n-1)/2}w_{j}(\zeta_{i(k)}, \epsilon_{i(k)})\in
L^{2} (\Sigma)\,;$$
then we have
\begin{eqnarray*}
\lefteqn{ \left\| \epsilon_{i(k)}^{-(n-1)/2}w_{j}(\zeta_{i(k)},
\epsilon_{i(k)})\right\|_{ L^{2}(S_{\epsilon_{i(k)}})} } \\ &&
\leq \left\| \epsilon_{i(k)}^{-(n-1)/2}w_{j}(\zeta_{i(k)},
\epsilon_{i(k)})-g \right\|_{L^{2}(\Sigma)} + \left(
\int_{S_{\epsilon_{i(k)}}} |g(x)|^{2}\,dS \right)^{1/2}\to 0
\end{eqnarray*}
as $k\to\infty$. Similarly we obtain
$$
 \left\| \epsilon_{i(k)}^{-(n-1)/2}
R(\overline{\zeta_{i(k)}},\epsilon_{i(k)}) w_{j}(\zeta_{i(k)},
\epsilon_{i(k)})\right\|_{ L^{2}(S_{\epsilon_{i(k)}})} \to
0\quad{\rm as}\quad k\to\infty.$$
Combining these result with the inequalities (\ref{west}) we infer
that
$$\epsilon_{i(k)}^{-(n-1)/2}\| w_{j}(\zeta_{i(k)},
\epsilon_{i(k)})\| _{H^{1}(\mathbb{R}^{n})}\to 0\quad\mathrm{as}
\quad k\to\infty\,, $$
which violates the relation (\ref{contr}); in this way we have
proved (\ref{wdecay2}).

Now we denote by $P_{\epsilon}$ the spectral projection of
$H_{\epsilon}$ associated with the interval $(\mu-3\kappa/4,
\mu+3\kappa/4)$. It follows from (\ref{wdecay2}) that
\begin{eqnarray*}
P_{\epsilon}\varphi_{j}-\varphi_{j} &\!=\!& \frac{\sqrt{-1}}{2\pi}
\oint_{|\zeta-\mu|=3\kappa/4}w_{j}(\zeta,\epsilon)\,d\zeta \\ &\!=\!&
o(\epsilon^{(n-1)/2}) \quad\mathrm{in} \quad H^{1}(\mathbb{R}^{n})
\quad\mathrm{as}\quad\epsilon\to 0
\end{eqnarray*}
holds for $m(\mu)\leq j\leq n(\mu)$. Consequently, we have
\begin{eqnarray}
\lefteqn{(H_{\epsilon}P_{\epsilon}\varphi_{i},P_{\epsilon}
\varphi_{j})_{L^{2}(\mathbb{R}^{n})}-\mu\delta_{i,j}
-\beta\varphi_{i}(0) \overline{\varphi_{j}(0)}\,
\mathrm{meas}_{\Sigma} (S_{\epsilon})} \nonumber
\\ && = q_{\epsilon}[P_{\epsilon}\varphi_{i},
P_{\epsilon}\varphi_{j}]-q_{0}[\varphi_{i},\varphi_{j}]
-\beta\varphi_{i}(0)\overline{\varphi_{j}(0)}\,
\mathrm{meas}_{\Sigma}(S_{\epsilon}) \nonumber\\ && =
q_{\epsilon}[\varphi_{i},\varphi_{j}]
-q_{0}[\varphi_{i},\varphi_{j}]-q_{\epsilon}[
(I\!-\!P_{\epsilon})\varphi_{i},(I\!-\!P_{\epsilon})\varphi_{j}]
\nonumber \\ && \phantom{A}
-\beta\varphi_{i}(0)\overline{\varphi_{j}(0)}
\,\mathrm{meas}_{\Sigma}(S_{\epsilon})\nonumber\\ && =
-q_{\epsilon}[(I\!-\!P_{\epsilon})\varphi_{i},
(I\!-\!P_{\epsilon})\varphi_{j}]
+\beta\int_{S_{\epsilon}}\varphi_{i}(x)\overline{
\varphi_{j}(x)}\,dS \nonumber \\ && \phantom{A}
-\beta\varphi_{i}(0)\overline{\varphi_{j}(0)}
\,\mathrm{meas}_{\Sigma}(S_{\epsilon})\nonumber\\ && =
o(\epsilon^{n-1}) \label{qconv}
\end{eqnarray}
and
\begin{equation} \label{Pconv}
(P_{\epsilon}\varphi_{i},P_{\epsilon}\varphi_{j})
_{L^{2}(\mathbb{R}^{n})} =\delta_{i,j} +o(\epsilon^{n-1})
\end{equation}
as $\epsilon\to 0$ for $m(\mu)\leq i,j\leq n(\mu)$, where we have
used, in the last step of
(\ref{qconv}),
 the assumptions (H.1), (H.2), the continuity of the restrictions
$\varphi_{i}|_{\Sigma}$ and $\varphi_{j}|_{\Sigma}$ at the origin,
 and the uniform boundedness of $q_{\epsilon}$ on $H^{1}(\mathbb{R}^{n})$
 with respect to $0<\epsilon\leq\eta_{0}$.
 Let us now introduce the matrices
\begin{eqnarray*}
L(\epsilon) &\!:=\!& ((H_{\epsilon}P_{\epsilon}\varphi_{i},
P_{\epsilon}\varphi_{j})_{L^{2}(\mathbb{R}^{n})}) _{m(\mu)\leq
i,j\leq n(\mu)}\,, \\ M(\epsilon) &\!:=\!&
((P_{\epsilon}\varphi_{i}, P_{\epsilon}\varphi_{j})
_{L^{2}(\mathbb{R}^{n})}) _{m(\mu)\leq i,j\leq n(\mu)}\,.
\end{eqnarray*}
Since $\{P_{\epsilon}\varphi_{j}\}_{m(\mu)\leq j\leq n(\mu)}$ is
 a basis of the spectral subspace ${\rm Ran}\, P_{\epsilon}$,
 we see that $\lambda_{m(\mu)}(\epsilon),
\lambda_{m(\mu)+1}(\epsilon),\ldots, \lambda_{n(\mu)}(\epsilon)$
are the eigenvalues of the matrix $L(\epsilon)M(\epsilon)^{-1}$,
which by (\ref{qconv}), (\ref{Pconv}) is equal to
$$ L(\epsilon)M(\epsilon)^{-1}=\mu I+\beta\,
\mathrm{meas}_{\Sigma}(S_{\epsilon})\, C(\mu)+o(\epsilon^{n-1}),$$
where $I$ stands for the identity matrix. This concludes the
argument. \QED


\section{Perturbation of an infinite curve}

As we have mentioned, the compactness of $\Sigma$ did not play an
essential role in the above argument, and we can use the same
technique for punctured noncompact manifolds of unit codimension
as well, as long as the corresponding Hamiltonian has a discrete
spectrum. At present this is known to be true in the case $n=2$
without restriction to the coupling constant $\beta$, see
\cite{EI}, and for $n=3$ and $\beta$ large enough \cite{EK2}.

We shall thus consider ``puncture" perturbations of infinite
asymptotically straight curves. Let $\Lambda:\, \mathbb{R}
\to\mathbb{R}^{2}$ be a $C^2$-smooth curve parameterized by its
arc length. Fix $\beta>0$ and assume that $\Lambda(0)=0$. Given
$\epsilon\geq 0$, we define
 $$ t_{\epsilon}[u,v]:= (\nabla u,\nabla
v)_{L^{2}(\mathbb{R}^{2})}-\beta \int_{\Lambda(\mathbb{R}\setminus
(-\epsilon,\epsilon))} u(x)\overline{v(x)}\,dS\,, \quad u,v\in
H^{1}(\mathbb{R}^{2})\,.$$
Let $T_{\epsilon}$ be the self-adjoint operator associated with
the quadratic form $t_{\epsilon}$. We adopt the following
assumptions about the curve $\Lambda$.
\begin{description}
 \vspace{-.6ex}
\item{(H.4)} The curve $\Lambda$ is not a straight line.
 \vspace{-1.2ex}
\item{(H.5)} There exists $c\in (0,1)$ such that
 $|\Lambda(s)-\Lambda
(t)|\geq c|t-s|$ for $s,t\in \mathbb{R}$.
 \vspace{-1.2ex}
\item{(H.6)} There exist $d>0$, $\rho>1/2$, and $w\in (0,1)$ such
 that the inequality
$$ 1-\frac {|\Lambda(s)-\Lambda(s^{\prime})|} {|s-s^{\prime}|}
\leq d\, \left\lbrack 1+|s+s^{\prime}|^{2\rho}
\right\rbrack^{-1/2} $$
holds in the sector $\left\{\, (s,s^{\prime})\in\mathbb{R}^{2};\:
w<\frac{s}{s^{\prime}}<w^{-1} \,\right\}$. \vspace{-.6ex}
\end{description}
From \cite[Prop~5.1 and Thm~5.2]{EI} we know that under these
conditions
$$\sigma_{\rm ess}(T_{0}) =[-\beta^{2}/4,\infty)\quad{\rm
and}\quad 1\leq\sharp\sigma_\mathrm{disc}(T_{0})\leq\infty.$$
Let $K:=\{j\in\mathbb{N};\: j\leq\sharp\sigma_\mathrm{disc}(T_{0})\}$.
 For $j\in K$, we denote by $\kappa_{j}(\epsilon)$ the $j$-th
eigenvalue of $T_{\epsilon}$ counted with multiplicity. The
function $\kappa_{j}(\cdot)$ is monotone non-decreasing,
continuous function in a neighbourhood of the origin. Let
$\{\psi_{j}(x)\}_{j\in K}$ be an orthonormal system of
eigenfunctions of $T_{0}$ such that $T_{0}\psi_{j}
=\kappa_{j}(0)\psi_{j}$ for $j\in K$. Each function
$\psi_{j}$ is continuous on $\Lambda$. For
$\mu\in\sigma_\mathrm{disc}(T_{0})$, we define
 \begin{eqnarray*}
p(\mu) &\!:=\!& \min\left\{\,j\in K;\: \mu=\kappa_{j}(0)
\,\right\}\,, \\ r(\mu) &\!:=\!& \max\left\{\,j\in K;\:
 \mu=\kappa_{j}(0)\,\right\}\,,
\\ D(\mu) &\!:=\!& \left(\psi_{i}(0) \overline{\psi_{j}(0)}
\right)_{p(\mu)\leq i,j\leq r(\mu)}\,.
 \end{eqnarray*}
Let $e_{p(\mu)}\leq e_{p(\mu)+1}\leq\cdots\leq e_{r(\mu)}$ be the
eigenvalues of the matrix $D(\mu)$. As in the compact case, if
$\mu=\kappa_j(0)$ is a simple eigenvalue of $H_0$, we have
$p(\mu)=r(\mu)=j$ and $e_j= |\psi_{j}(0)|^2$. The asymptotic
behaviour now looks as follows.

\begin{theorem} \label{curve}
Assume that (H.4)--(H.6) and take $\mu\in
\sigma_\mathrm{disc}(T_{0})$. Then
$$ \kappa_{j}(\epsilon)=\mu +2\beta e_{j}\epsilon
+o(\epsilon)\quad\mathit{as}\quad\epsilon\to 0 $$
holds for $p(\mu)\leq j\leq r(\mu)$.
\end{theorem}
{\sl Proof} is analogous to that of Theorem~\ref{main}. \QED
\vspace{1em}

Let us mention in conclusion that the results derived here raise
some interesting questions, for instance, what is the following
term in the expansion, what the asymptotic behaviour looks like for
non-smooth surfaces, and whether similar formulae are valid in the
case of $\mathrm{codim\,}\Sigma=2,3$ when the corresponding
generalized Schr\"odinger operator has to be defined by means of
appropriate boundary conditions.


\subsection*{Acknowledments}

The authors are grateful for the hospitality extended to them,
P.E. in the Department of Mathematics, Tokyo Metropolitan
University, and K.Y. in the Nuclear Physics Institute, AS CR;
during these visits a part of this work was done. The research has
been partially supported by GAAS and the Czech Ministry of
Education within the projects A1048101 and ME482. Useful comments
by the referees are also appreciated.

\end{document}